\def\be{\begin{equation}}
\def\ee{\end{equation}}
\def\ba{\begin{array}}
\def\ea{\end{array}}

\documentclass[aps,amsmath,amssymb,amsfonts,showpacs]{revtex4}
\usepackage{graphicx}
\usepackage{epstopdf}
\def\qed{\leavevmode\unskip\penalty9999 \hbox{}\nobreak\hfill
     \quad\hbox{\leavevmode  \hbox to.77778em{%
               \hfil\vrule   \vbox to.675em%
               {\hrule width.6em\vfil\hrule}\vrule\hfil}}
     \par\vskip3pt}

\newtheorem{theorem}{Theorem}

\newtheorem{lemma}{Lemma}

\begin{document}

\title{Tighter monogamy and polygamy relations using R\'{e}nyi-$\alpha$ entropy}

\author{Yanying Liang$^1$, Zhu-Jun Zheng$^1$, Chuan-Jie Zhu$^2$}

\affiliation{$^1$School of Mathematics, South China University of Technology, Guangzhou 510641, China\\
 $^2$Department of Physics, Renmin University of China,   Beijing 100872,  China}

\begin{abstract}

We investigate monogamy relations related to the R\'{e}nyi-$\alpha$ entanglement and  polygamy relations related to the R\'{e}nyi-$\alpha$ entanglement of assistance. We present new entanglement monogamy relations satisfied by the $\mu$-th power of  R\'{e}nyi-$\alpha$ entanglement with $\alpha\in[(\sqrt{7}-1)/2,(\sqrt{13}-1)/2]$ for $\mu\geqslant2$, and polygamy relations satisfied by the $\mu$-th power of R\'{e}nyi-$\alpha$ entanglement of assistance with $\alpha\in[(\sqrt{7}-1)/2,(\sqrt{13}-1)/2]$ for $0\leq\mu\leq1$. These relations are shown to be tighter than the existing ones.

\end{abstract}


\maketitle

\section{Introduction}

One fundamental property of quantum entanglement is in its limited shareability in multi-party
quantum systems \cite{hor09rmp}. For example, if the two subsystems are more entangled with each other, then they will share a less amount of entanglement with the other subsystems with specific entanglement measures. This restricted shareability of entanglement is named as the monogamy of entanglement (MoE). The concept of monogamy is an essential feature allowing
for security in quantum key distribution \cite{Pawlowski10pra}. It also
plays an important role in many field of physics such
as foundations of quantum mechanics \cite{Bennett14,Toner09,Seevinck10qip}, condensed matter
physics \cite{Ma11np,Saez13prb}, statistical mechanics \cite{Bennett14}, and
even black-hole physics \cite{Susskind13,Lloyd14}. Monogamy inequality was first built for three-qubit
systems using tangle as the bipartite entanglement measure \cite{ckw00pra}, and generalized into multi-qubit systems in terms of various entanglement measures \cite{osb06prl}.

On the other hand, the assisted entanglement, which is a dual concept to bipartite
entanglement measures, is known to have a dually monogamous or polygamous property in multiparty quantum systems. The polygamous property can be regarded as another kind of entanglement constraints in multi-qubit systems, and Gour \emph{et al} \cite{Gour05pra} established the first dual monogamy inequality or polygamy inequality for multi-qubit systems using concurrence
of assistance (CoA). For a three-qubit pure state ${\left| \psi  \right\rangle _{A_1 A_2 A_3 } }$, a polygamy inequality was introduced as:
\begin{eqnarray}\label{q1}
C^2 \left( {\left| \psi  \right\rangle _{A_1 |A_2 A_3 } } \right) \le \left[ {C^a \left( {\rho _{A_1 A_2 } } \right)} \right]^2  + \left[ {C^a \left( {\rho _{A_1 A_3 } } \right)} \right]^2,
\end{eqnarray}
where CoA for a bipartite state ${\rho _{AB} }$ is defined as: $C^a \left( {\rho _{AB} } \right) =  \max_{\{p_{i}, |\psi_{i}\rangle\}}\sum_{i}p_{i}C(|\psi_{i}\rangle_{AB})$, with the maximum is taken over all possible pure state decompositions of ${\rho _{AB} }= \sum_{i} {p_i \left| {\psi _i } \right\rangle _{AB} \left\langle {\psi _i } \right|}$ and ${C\left( {\left| {\psi _i } \right\rangle _{AB} } \right)}$ denotes the concurrence \cite{woo98prl} of ${\left| {\psi _i } \right\rangle _{AB} }$. Furthermore, it is shown that for any pure state $\left| \psi  \right\rangle _{A_1 A_2  \cdots A_n }$ in a $n$-qubit system \cite{Gour07jmp}, we have
\begin{eqnarray}\label{q2}
C^2 \left( {\left| \psi  \right\rangle _{A_1 |A_2  \cdots A_n } } \right) \le \left[ {C^a \left( {\rho _{A_1 A_2 } } \right)} \right]^2  +  \cdots  + \left[ {C^a \left( {\rho _{A_1 A_n } } \right)} \right]^2.
\end{eqnarray}

R\'{e}nyi-$\alpha$ entanglement (R$\alpha$E) \cite{hor96pla} is a well-defined entanglement measure
which is the generalization of entanglement of formation (EOF) and has the merits for characterizing
quantum phases with differing computational power \cite{jcui12nc}, ground state properties in many-body
systems \cite{fran14prx}, and topologically ordered states \cite{fla09prl,hal13prl}. Therefore, it is natural to study the monogamy inequality of  R$\alpha$E and the polygamy inequality of  R$\alpha$E of assistance in multipartite entanglement detection.

In this paper, we show that the monogamy inequality of R$\alpha$E and the polygamy inequality of  R$\alpha$E of assistance obtained so far can be made tighter. When $\alpha\in[(\sqrt{7}-1)/2,(\sqrt{13}-1)/2]$, we establish entanglement monogamy relations for the $\mu$-th power of  R$\alpha$E with $\mu\geq2$ and polygamy relations for the $\mu$-th power of R$\alpha$E of assistance with $0\leq\mu\leq1$ which are tighter than those in \cite{WS,WS2}.

\section{Tighter monogamy relations for R\'{e}nyi-$\alpha$ entanglement}

Let $\mathbf{H}_X$ denote a discrete finite-dimensional complex vector space associated with a quantum subsystem $X$. For a bipartite pure state $|\psi\rangle_{AB}$ in vector space $\mathbf{H}_A\otimes \mathbf{H}_B$, the R$\alpha$E is defined as
\cite{JSK4}
\begin{equation}\label{q3}
E_{\alpha}(\left| \psi  \right\rangle _{AB})= S_\alpha(\rho_A) = \frac{1}{1-\alpha}\log _2(\mbox{tr}\rho _A^\alpha),
\end{equation}
where the R\'{e}nyi-$\alpha$ entropy is $S_\alpha(\rho_A)=[\log _2(\sum_i\lambda_i^{\alpha})]/(1-\alpha)$ with
$\alpha$ being a nonnegative real number and $\lambda_i$ being the eigenvalue of reduced density matrix
$\rho_A$. The R\'{e}nyi-$\alpha$ entropy $S_\alpha \left( \rho  \right)$ converges to the von Neumann
entropy when the order $\alpha$ tends to 1. For a bipartite mixed state $\rho _{AB}$, the R$\alpha$E
is defined via the convex-roof extension
\begin{eqnarray}\label{q4}
E_\alpha(\rho _{AB})=\min \sum_i p_i E_\alpha(|{\psi _i }\rangle_{AB}),
\end{eqnarray}
where the minimum is taken over all possible pure state decompositions of ${\rho_{AB}=\sum_{i}{p_i
\left| {\psi _i } \right\rangle _{AB} \left\langle {\psi _i } \right|} }$.

In particular, for a bipartite $2 \otimes d$ mixed state $\rho _{AB}$, the
R\'{e}nyi-$\alpha$ entanglement has an analytical expression \cite{WS}
\begin{eqnarray}\label{q12}
E_\alpha  \left( {\rho _{AB} } \right) = f_\alpha  \left[ {C^2 \left( {\rho _{AB} } \right)} \right],
\end{eqnarray}
where the order $\alpha$ ranges in the region $[(\sqrt 7  - 1)/2,(\sqrt {13}  - 1)/2]$ and the function $f_\alpha \left( x \right)$ has the form
\begin{equation}\label{q6}
f_\alpha \! \left( x \right)\!= \!\frac{1}{{1 - \alpha }}\!\log _2 \!\left[ {\left( {\frac{{1 \!-\!
\sqrt {1 - x} }}{2}} \right)^\alpha  \!\!\!\! +\! \left( {\frac{{1 \!+\! \sqrt {1 - x} }}{2}}
\right)^\alpha  } \right].
\end{equation}

For any
two-qubit state ${\rho _{AB} }$ with $\alpha  \ge \left( {\sqrt 7  - 1} \right)/2$, there also exist an analytic formula of R$\alpha$E \cite{JSK4}
\begin{eqnarray}\label{q5}
E_\alpha  \left( {\rho _{AB} } \right) = f_\alpha  \left[ {C\left( {\rho _{AB} } \right)} \right],
\end{eqnarray}
where the function $f_\alpha \left( x \right)$ has the form (6).

In Ref.\cite{WS}, we have known that for an arbitrary three-qubit mixed state ${\rho _{A_1 A_2 A_3 } }$, the $\mu$-th power
R\'{e}nyi-$\alpha$ entanglement obeys the monogamy relation
\begin{eqnarray}\label{q25}
E_\alpha^\mu  \left( {\rho _{A_1 |A_2 A_3 } } \right) \ge E_\alpha^\mu  \left({\rho _{A_1 A_2}}\right)
+ E_\alpha^\mu  \left( {\rho _{A_1 A_3 } } \right),
\end{eqnarray}
where the order $\alpha\ge(\sqrt{7}-1)/2\simeq 0.823$ and the power $\mu \geq 2$. Moreover, in $N$-qubit
systems, the following monogamy relation is also satisfied
\begin{eqnarray}\label{q26}
E_\alpha ^\mu  ( {\rho _{A|B_1B_2\cdots B_{N-1} } } ) \ge \sum_{i = 1}^{k - 1} {E_\alpha ^\mu
({\rho _{AB_i }})}+E_{\alpha}^{\mu}(\rho_{A|B_k\dots B_{N-1}}),
\end{eqnarray}
where the power $\mu\geq 2$ and the order $\alpha\in[(\sqrt{7}-1)/2,(\sqrt{13}-1)/2]$.

In fact, we can prove the following results for  R\'{e}nyi-$\alpha$ entanglement. Before this, we need to consider a Lemma for concurrence.

\begin{lemma} \cite{1803} For any $2\otimes2\otimes2^{n-2}$ mixed state $\rho\in \mathbf{H}_A\otimes \mathbf{H}_{B}\otimes \mathbf{H}_{C}$, if $C_{AB}\geqslant C_{AC}$, we have
\begin{equation}\label{le2}
  C^\alpha_{A|BC}\geqslant  C^\alpha_{AB}+(2^{\frac{\alpha}{2}}-1)C^\alpha_{AC},
\end{equation}
for all $\alpha\geqslant2$.
\end{lemma}

[Proof]
Since it has been shown that $C^2_{A|BC}\geqslant C^2_{AB}+C^2_{AC}$ for arbitrary $2\otimes2\otimes2^{n-2}$ tripartite state $\rho_{ABC}$ \cite{11}. Then, if $C_{AB}\geqslant C_{AC}$, we have
\begin{eqnarray*}
  C^\alpha_{A|BC}&&\geqslant (C^2_{AB}+C^2_{AC})^{\frac{\alpha}{2}}\\
  &&=C^\alpha_{AB}\left(1+\frac{C^2_{AC}}{C^2_{AB}}\right)^{\frac{\alpha}{2}} \\
  && \geqslant C^\alpha_{AB}\left[1+(2^{\frac{\alpha}{2}}-1)\left(\frac{C^2_{AC}}{C^2_{AB}}\right)^{\frac{\alpha}{2}}\right]\\
  &&=C^\alpha_{AB}+(2^{\frac{\alpha}{2}}-1)C^\alpha_{AC}
\end{eqnarray*}
where the second inequality is due to $(1+t)^x\geqslant 1+(2^{x}-1)t^x$ for any real number $x$ and $t$, $0\leqslant t \leqslant 1$, $x\in [1, \infty]$. As the subsystems $A$ and $B$ are equivalent in this case, we have assumed that $C_{AB}\geqslant C_{AC}$ without loss of generality. Moreover,
if $C_{AB}=0$ we have $C_{AB}=C_{AC}=0$. That is to say the lower bound becomes trivially zero.
\qed

\begin{theorem}
 For any $N$-qubit mixed state $\rho\in \mathbf{H}_A\otimes \mathbf{H}_{B_1}\otimes\cdots\otimes \mathbf{H}_{{B_{N-1}}}$, if
${C_{AB_i}}\geqslant {C_{A|B_{i+1}\cdots B_{N-1}}}$ for $i=1, 2, \cdots, m$, and
${C_{AB_j}}\leqslant {C_{A|B_{j+1}\cdots B_{N-1}}}$ for $j=m+1,\cdots,N-2$, $\forall$ $1\leqslant m\leqslant N-3$, $N\geqslant 4$, the R\'{e}nyi-$\alpha$ entanglement $E_\alpha(\rho)$ satisfies
\begin{eqnarray}
&&~~~E_\alpha ^\mu  ( {\rho _{A|B_1B_2\cdots B_{N-1} } } )\nonumber\\
&&~~~~~~\geqslant E_\alpha ^\mu({\rho _{AB_1 }})+(2^{\mu}-1)E_\alpha ^\mu({\rho _{AB_2 }})+\cdots+(2^{\mu}-1)^{m-1}E_\alpha ^\mu({\rho _{AB_m }})\nonumber\\
&&~~~~~~~~~+(2^{\mu}-1)^{m+1}\left(E_\alpha ^\mu({\rho _{AB_{m+1} }})+\cdots+E_\alpha ^\mu({\rho _{AB_{N-2} }})\right)\nonumber\\
&&~~~~~~~~~+(2^{\mu}-1)^{m}E_\alpha ^\mu({\rho _{AB_{N-1} }}),
\end{eqnarray}
for $\mu\geqslant2$ and $\alpha\in[(\sqrt{7}-1)/2,(\sqrt{13}-1)/2]$.
\end{theorem}

 [Proof]
 For $\mu\geqslant2$, we have
\begin{eqnarray}\label{FA}
 f_\alpha^{{\mu}}(x^2+y^2)&&\geqslant  \left(f_\alpha(x^2)+f_\alpha(y^2)\right)^{\mu}\nonumber\\
 &&\geqslant  f_\alpha^{\mu}(x^2)+(2^{\mu}-1)f_\alpha^{\mu}(y^2),
\end{eqnarray}
where the first inequality is due to the convex property of $f_\alpha(x)$ for $\alpha\ge(\sqrt{7}-1)/2$ \cite{JSK4}, and the second inequality is obtained from a similar consideration in the proof of the second inequality in Lemma 1.

Let $\rho=\sum_ip_i|\psi_i\rangle\langle\psi_i|\in \mathbf{H}_A\otimes \mathbf{H}_{B_1}\otimes\cdots\otimes \mathbf{H}_{{B_N-1}}$ be the optimal decomposition of $E_\alpha({\rho _{A|B_1B_2\cdots B_{N-1} } })$ for the $N$-qubit mixed state $\rho$; then we have
\begin{eqnarray}
E_\alpha^2({\rho _{A|B_1B_2\cdots B_{N-1} } })&=&[\sum_i p_i E_\alpha(|{\psi_i}\rangle_{A|B_1B_2\cdots B_{N-1} })]^2\nonumber\\
&=&\{\sum_i p_i E_\alpha[C_{A|B_1B_2\cdots B_{N-1} }(|{\psi_i}\rangle)]\}^2\nonumber\\
&\geq& \{E_\alpha[\sum_i p_i C_{A|B_1B_2\cdots B_{N-1} }(|{\psi_i}\rangle)]\}^2\nonumber\\
&\geq& \{E_\alpha[C_{A|B_1B_2\cdots B_{N-1} }(\rho)]\}^2\nonumber\\
&=&E_\alpha^2[C^2_{A|B_1B_2\cdots B_{N-1} }(\rho)],
\end{eqnarray}
here we have used in the second equality the pure state formula of the R$\alpha$E and taken the
$E_\alpha(C)$ as a function of the concurrence $C$ for $\alpha \ge (\sqrt {7} - 1)/2$; in the
third inequality we have used the monotonically increasing and convex properties of $E_\alpha(C)$ as a
function of the concurrence \cite{JSK4}; in the forth inequality we have used the convex property
of concurrence for mixed states. Then from (13) we have
\begin{eqnarray}
   E_\alpha ^\mu  ( {\rho _{A|B_1B_2\cdots B_{N-1} } } )
&& \geqslant f_\alpha^\mu(C^2_{AB_1}+C^2_{AB_2}+\cdots+C^2_{AB_{m-1}})\nonumber\\
&& \geqslant f_\alpha^\mu(C^2_{AB_1})+(2^{\mu}-1) f_\alpha^\mu(C^2_{AB_2}+\cdots+C^2_{AB_{m-1}})\nonumber\\
&& \geqslant f_\alpha^\mu(C^2_{AB_1})+(2^{\mu}-1) f_\alpha^\mu(C^2_{AB_2})+(2^{\mu}-1)^2f_\alpha^\mu(C^2_{AB_3}+\cdots+C^2_{AB_{m-1}})\nonumber\\
&&  \geqslant \cdots\nonumber\\
&&  \geqslant f_\alpha^\mu(C^2_{AB_1})+(2^{\mu}-1) f_\alpha^\mu(C^2_{AB_2})+\cdots+(2^{\mu}-1)^{m-1} f_\alpha^\mu(C^2_{AB_m})\nonumber\\
&&                            +(2^{\mu}-1)^{m}f_\alpha^\mu(C^2_{A|B_{m+1}\cdots B_{N-1}}),
\end{eqnarray}
\normalsize
where we have used the monogamy inequality $C^x(\rho_{A|B_1B_2 \cdots B_{N-1}})\geqslant C^x(\rho_{AB_1})+C^x(\rho_{AB_2})+\cdots+C^x(\rho_{AB_{N-1}})$ with $x\geq2$ for $N$-qubit states $\rho$ and the monotonically increasing property of $f_\alpha(C^2)$ to obtain the first inequality. By using (12) repeatedly, we get the other inequalities.

Since ${C_{AB_i}}\geqslant {C_{A|B_{i+1}\cdots B_{N-1}}}$ for $i=1, 2, \cdots, m$, and
${C_{AB_j}}\leqslant {C_{A|B_{j+1}\cdots B_{N-1}}}$ for $j=m+1,\cdots,N-2$, $\forall$ $1\leqslant m\leqslant N-3$, $N\geqslant 4$, by using (12) and the similar consideration in the proof of the second inequality in Lemma 1, then we have
\begin{eqnarray}
  f_\alpha^\mu(C^2_{A|B_{m+1}\cdots B_{N-1}})
&& \geqslant (2^{\mu}-1)f_\alpha^\mu(C^2_{AB_{m+1}})+f_\alpha^\mu(C^2_{A|B_{m+2}\cdots B_{N-1}})\nonumber\\
&& \geqslant (2^{\mu}-1)\left(f_\alpha^\mu(C^2_{AB_{m+1}})+\cdots+f_\alpha^\mu(C^2_{AB_{N-2}})\right)\nonumber\\
&&            +f_\alpha^\mu(C^2_{AB_{N-1}}).
\end{eqnarray}
\normalsize

Since for any $2\otimes2$ quantum state $\rho_{AB_i}$, $\alpha\in[(\sqrt{7}-1)/2,(\sqrt{13}-1)/2]$, $E_\alpha(\rho_{AB_i})=f_\alpha\left[C^2(\rho_{AB_i})\right]$, therefore combining (14) and (15), we have Theorem 1.
\qed

Moreover, for the case that ${C_{AB_i}}\geqslant {C_{A|B_{i+1}\cdots B_{N-1}}}$ for all $i=1, 2, \cdots, N-2$, we have a simple tighter monogamy relation for the R\'{e}nyi-$\alpha$ entanglement:

\begin{theorem}
If ${C_{AB_i}}\geqslant {C_{A|B_{i+1}\cdots B_{N-1}}}$ for all $i=1, 2, \cdots, N-2$, we have
\begin{eqnarray}\label{th2}
&&~~~E_\alpha ^\mu  ( {\rho _{A|B_1B_2\cdots B_{N-1} } } )\nonumber\\
&&~~~~~~\geqslant E_\alpha ^\mu({\rho _{AB_1 }})+(2^{\mu}-1)E_\alpha ^\mu({\rho _{AB_2 }})\cdots+(2^{\mu}-1)^{N-2}E_\alpha ^\mu({\rho _{AB_{N-1} }}),
\end{eqnarray}
for $\mu\geqslant2$ and $\alpha\in[(\sqrt{7}-1)/2,(\sqrt{13}-1)/2]$.
\end{theorem}

\begin{figure}
  \centering
  \includegraphics[width=7cm]{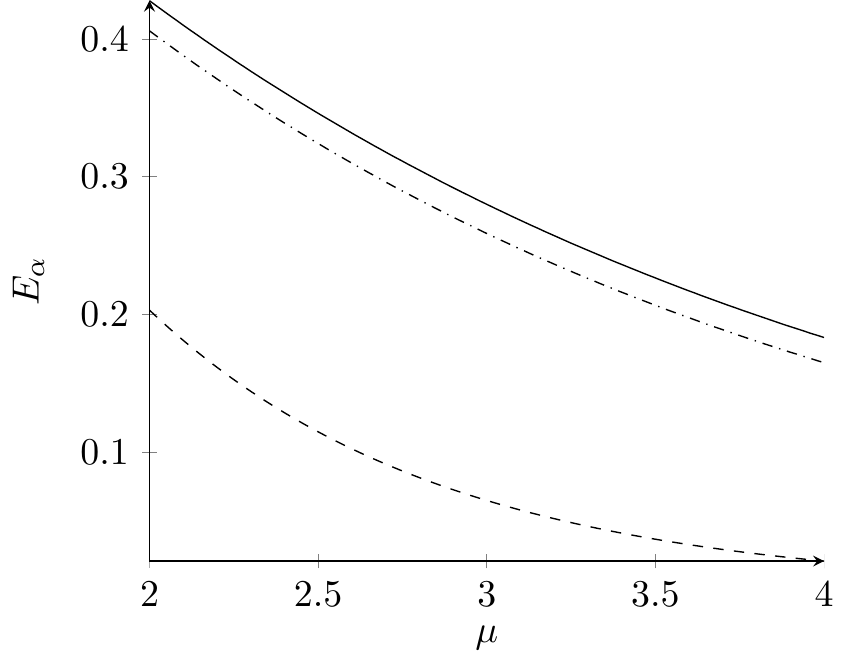}\\
  \caption{Behavior of the R\'{e}nyi-$\alpha$ entanglement of $|\psi\rangle$ and its lower bound, which are functions of $\mu$ plotted. The solid line represents the R\'{e}nyi-$\alpha$ entanglement of $|\psi\rangle$ in Example 1, the dot-dashed line represents the lower bound from our result, and the dashed line represents the lower bound from the result in (9) of \cite{WS}.}
\end{figure}

As an example, let us consider the three-qubit state $|\psi\rangle$ in the generalized Schmidt decomposition form \cite{38,39},
\begin{eqnarray}\label{ex1}
|\psi\rangle&=&\lambda_0|000\rangle+\lambda_1e^{i{\varphi}}|100\rangle+\lambda_2|101\rangle \nonumber\\
&&+\lambda_3|110\rangle+\lambda_4|111\rangle,
\end{eqnarray}
where $\lambda_i\geqslant0,~i=0,1,2,3,4$ and $ \sum_{i = 0}^{4} \lambda_i^2=1.$ Set $\lambda_{0}=\lambda_{1}=\frac{1}{2}$, $\lambda_{2}=\lambda_{3}=\lambda_{4}=\frac{\sqrt{6}}{6}$. Since $\alpha\in[(\sqrt{7}-1)/2,(\sqrt{13}-1)/2]$, we choose $\alpha=(\sqrt{7}-1)/2\approx0.823$, we have
$E_\alpha({|\psi\rangle_{A|B}})=E_\alpha({|\psi\rangle_{A|C
}})=0.318620$, $E_\alpha({|\psi\rangle_{A|BC}})=0.654205$, and then $E_\alpha^{\mu}({|\psi\rangle_{A|BC}})=(0.654205)^{\mu}$, $E_\alpha^{\mu}({|\psi\rangle_{A|B}})+E_\alpha^{\mu}({|\psi\rangle_{A|C}})=2(0.318620)^{\mu}$, $E_\alpha^{\mu}({|\psi\rangle_{A|B}})+(2^{\mu}-1)E_\alpha^{\mu}({|\psi\rangle_{A|C}})=2^{\mu}(0.318620)^{\mu}$. It is easily verified that our result is better than the result in (9) for $\mu\geqslant2$; see Fig 1.

\section{Tighter polygamy relations for  R\'{e}nyi-$\alpha$ entanglement of assistance}

As a dual concept to R\'{e}nyi-$\alpha$ entanglement, we define the R\'{e}nyi-$\alpha$ entanglement of assistance (REoA) as
\begin{eqnarray}\label{q7}
E_\alpha ^a \left( {\rho _{AB} } \right) = \max \sum_i {p_i } E_\alpha  \left( {\left| {\psi _i } \right\rangle _{AB} } \right),
\end{eqnarray}
where the maximum is taken over all possible pure state decompositions of ${\rho _{AB} }= \sum\nolimits_i {p_i \left| {\psi _i } \right\rangle _{AB} \left\langle {\psi _i } \right|}$.

In Ref. \cite{WS2}, we know that for any two-qubit state $\rho _{AB}$ and $\alpha  \ge \left( {\sqrt 7  - 1} \right)/2$, we have
\begin{eqnarray}\label{q8}
E_\alpha^a \left( {\rho _{AB} } \right) \ge f_{\alpha}\left( {C^a \left( {\rho _{AB} } \right)} \right),
\end{eqnarray}
where $E_\alpha^a \left( {\rho _{AB} } \right)$ and ${C^a \left( {\rho _{AB} } \right)}$ are the REoA and CoA of ${\rho _{AB} }$, respectively.
And for any $\left( {\sqrt 7  - 1} \right)/2 \le \alpha  \le \left( {\sqrt {13}  - 1} \right)/2$ and the function $f_\alpha  \left( x \right)$ defined on the domain $\mathcal{D} = \left\{ {\left( {x, y} \right)|0 \leq x,y \leq 1, 0 \leq x^2+y^2 \leq 1} \right\}$, we have
\begin{eqnarray}\label{q10}
f_\alpha  (\sqrt {x^2  + y^2 } ) \le f_\alpha  (x) + f_\alpha  (y).
\end{eqnarray}

From Ref. \cite{WS2}, it has been shown that for $\left( {\sqrt {7}  - 1} \right)/2\leq \alpha \leq \left( {\sqrt {13}  - 1}\right)/2, 0\leq\mu\leq1$, and any $N$-qubit state $\rho _{A|B_1B_2\cdots B_{N-1} } $, we have
\begin{eqnarray}\label{q24}
\left[ {E_\alpha ^a \left( {\rho _{A|B_1B_2\cdots B_{N-1} }} \right)} \right]^\mu   \le \left[ {E_\alpha ^a \left( {\rho _{A|B_1}}  \right)} \right]^\mu   +  \cdots  + \left[ {E_\alpha ^a \left( {\rho _{A|B_{N-1} }} \right)} \right]^\mu.
\end{eqnarray}

In the following, we study the polygamy relations of  REoA for $N$-qubit generalized $W$-class state.
For $N$-qubit generalized $W$-class state, $|\psi\rangle_{AB_1\cdots B_{N-1}}\in H_A\otimes H_{B_1}\otimes\cdots\otimes H_{B_{N-1}}$ defined by
\begin{eqnarray}\label{gw}
|\psi\rangle_{AB_1\cdots B_{N-1}}=a|10\cdots0\rangle+b_1|01\cdots0\rangle+\cdots+b_{N-1}|00\cdots1\rangle,
\end{eqnarray}
with $|a|^2+\sum_{i=1}^{N-1}|b_i|^2=1$,
one has \cite{16},
\begin{eqnarray}\label{la2}
C(\rho_{AB_i})=C^a(\rho_{AB_i}),~~~~i=1,2,...,N-1,
\end{eqnarray}
where $\rho_{AB_i}=Tr_{B_1\cdots B_{i-1}B_{i+1}\cdots B_{N-1}}(|\psi\rangle_{AB_1\cdots B_{N-1}}\langle\psi|)$.

\begin{theorem}
Let $\rho_{AB_1\cdots B_{N-1}}$ denote the $N$-qubit reduced density matrix of the $N$-qubit generalized $W$-class state $|\psi\rangle_{AB_1\cdots B_{N-1}}\in H_A\otimes H_{B_1}\otimes\cdots\otimes H_{B_{N-1}}$, if
${C_{AB_i}}\geqslant {C_{A|B_{i+1}\cdots B_{N-1}}}$ for $i=1, 2, \cdots, m$, and
${C_{AB_j}}\leqslant {C_{A|B_{j+1}\cdots B_{N-1}}}$ for $j=m+1,\cdots,N-2$, $\forall$ $1\leqslant m\leqslant N-3$, $N\geqslant 4$, the R\'{e}nyi-$\alpha$ entanglement of assistance $E_\alpha^a(\rho)$ satisfies
\begin{eqnarray}
&&~~~\left[ {E_\alpha ^a \left( {\rho _{A|B_1B_2\cdots B_{N-1} }} \right)} \right]^\mu  \nonumber\\
&&~~~~~~\leqslant \left[ {E_\alpha ^a \left( {\rho _{A|B_1}}  \right)} \right]^\mu +(2^{\mu}-1)\left[ {E_\alpha ^a \left( {\rho _{A|B_2}}  \right)} \right]^\mu+ \cdots+(2^{\mu}-1)^{m-1}\left[ {E_\alpha ^a \left( {\rho _{A|B_m}}  \right)} \right]^\mu \nonumber\\
&&~~~~~~~~~+(2^{\mu}-1)^{m+1}\left(\left[ {E_\alpha ^a \left( {\rho _{A|B_{m+1}}}  \right)} \right]^\mu +\cdots+\left[ {E_\alpha ^a \left( {\rho _{A|B_{N-2}}}  \right)} \right]^\mu \right)\nonumber\\
&&~~~~~~~~~+(2^{\mu}-1)^{m}\left[ {E_\alpha ^a \left( {\rho _{A|B_{N-1}}}  \right)} \right]^\mu ,
\end{eqnarray}
for $0\leq\mu\leq1$ and $\alpha\in[(\sqrt{7}-1)/2,(\sqrt{13}-1)/2]$.
\end{theorem}

[Proof]
For $0\leq\mu\leq1$, we have
\begin{eqnarray}
 \left[f_\alpha(\sqrt {x^2+ y^2})\right]^{\mu} &&\le \left[f_\alpha(x) + f_\alpha(y)\right]^{\mu}\nonumber\\
 && \le  f_\alpha^{\mu}(x)+(2^{\mu}-1)f_\alpha^{\mu}(y),
\end{eqnarray}
where the first inequality is due to inequality (20) and the monotonically increasing property of $x^\mu$ for $0\leq\mu\leq1$, and the second equality is obtained from a similar consideration in the proof of the second inequality in Lemma 1. Here we note that $(1+t)^x\leqslant 1+(2^{x}-1)t^x$ with $0\leqslant t \leqslant 1$, $x\in [0, 1]$.

For the $N$-qubit generalized $W$-class state $\left|\psi\right\rangle _{A|B_1B_2\cdots B_{N-1}}$, from Eq.(2), we have
\begin{eqnarray}
C^2\left({\left|\psi\right\rangle _{A|B_1B_2\cdots B_{N-1} }} \right) \le \left[ {C^a \left({\rho _{A|B_1 }} \right)} \right]^2  +  \cdots  + \left[ {C^a \left( {\rho _{A|B_{N-1}} } \right)} \right]^2.
\end{eqnarray}
Assuming that $ {C^2\left( {\rho _{A|B_1B_2\cdots B_{N-1} }} \right)}   \le \left[ {C^a \left( {\rho _{A|B_1}  } \right)} \right]^2  +  \cdots  + \left[ {C^a \left( {\rho _{A|B_{N-1}}  } \right)} \right]^2  \le 1$ , then
\begin{eqnarray}
&&~~~\left[ {E_\alpha ^a \left( { \rho _{A|B_1B_2\cdots B_{N-1} }} \right)} \right]^\mu \nonumber\\
  && = f_\alpha^\mu \left( {C\left( {\rho _{A|B_1B_2\cdots B_{N-1} } } \right)} \right)\nonumber\\
  && \le
  f_\alpha^\mu  \left( {\sqrt {\left[ {C^a \left( {\rho _{A|B_1} } \right)} \right]^2  +  \cdots  + \left[ {C^a \left( {\rho _{A|B_{N-1}} } \right)} \right]^2 } } \right) \nonumber\\
  && = f_\alpha^\mu  \left( {\sqrt {\left[ {C \left( {\rho _{A|B_1} } \right)} \right]^2  +  \cdots  + \left[ {C \left( {\rho _{A|B_{N-1}} } \right)} \right]^2 } } \right) \nonumber\\
  && \le
  f_\alpha^\mu \left( {C \left( {\rho _{A|B_1} } \right)} \right) +(2^{\mu}-1)f_\alpha^\mu \left( {C \left( {\rho _{A|B_2} } \right)} \right)+ \cdots+(2^{\mu}-1)^{m-1}f_\alpha^\mu \left( {C \left( {\rho _{A|B_m} } \right)} \right) \nonumber\\
  && +(2^{\mu}-1)^{m+1}\left(f_\alpha^\mu \left( {C \left( {\rho _{A|B_{m+1}} } \right)} \right)+\cdots+f_\alpha^\mu \left( {C \left( {\rho _{A|B_{N-2}} } \right)} \right) \right)\nonumber\\
  && +(2^{\mu}-1)^{m} {f_\alpha ^\mu \left( C \left({\rho _{A|B_{N-1}}} \right) \right)} \nonumber\\
  && \le
  \left[ {E_\alpha ^a \left( {\rho _{A|B_1}}  \right)} \right]^\mu +(2^{\mu}-1)\left[ {E_\alpha ^a \left( {\rho _{A|B_2}}  \right)} \right]^\mu \cdots+(2^{\mu}-1)^{m-1}\left[ {E_\alpha ^a \left( {\rho _{A|B_m}}  \right)} \right]^\mu \nonumber\\
  && +(2^{\mu}-1)^{m+1}\left(\left[ {E_\alpha ^a \left( {\rho _{A|B_{m+1}}}  \right)} \right]^\mu +\cdots+\left[ {E_\alpha ^a \left( {\rho _{A|B_{N-2}}}  \right)} \right]^\mu \right)\nonumber\\
  && +(2^{\mu}-1)^{m}\left[ {E_\alpha ^a \left( {\rho _{A|B_{N-1}}}  \right)} \right]^\mu ,
\end{eqnarray}
where in the second inequality we have used the monotonically increasing property of $f_\alpha(x)$ for $\alpha \ge \left( {\sqrt 7  - 1} \right)/2$, and the third equality is due to (23). By using (25) repeatedly and the similar consideration in the proof of Theorem 1, we get the forth inequality. The last inequality is due to (19) and (23).

Then we consider the case $ {C^2\left( {\rho _{A|B_1B_2\cdots B_{N-1} }} \right)}    \le 1 \le \left[ {C^a \left( {\rho _{A|B_1 } } \right)} \right]^2  +  \cdots  + \left[ {C^a \left( {\rho _{A|B_{N-1} } } \right)} \right]^2$. There must exist $k \in \left\{ {1, \ldots ,N - 2} \right\}$ such that  $\left[ {C^a \left( {\rho _{A|B_1 }} \right)} \right]^2  +  \cdots  + \left[ {C^a \left( {\rho _{A|B_k } } \right)} \right]^2  \le 1, \left[ {C^a \left( {\rho _{A|B_1 } } \right)} \right]^2  +  \cdots  + \left[ {C^a \left( {\rho _{A|B_{k+1} }}\right)} \right]^2  > 1$. By defining $T = \left[ {C^a \left( {\rho _{A|B_1 }} \right)} \right]^2  +  \cdots  + \left[ {C^a \left( {\rho _{A|B_{k+1} } } \right)} \right]^2  - 1 > 0$,
we can derive
\begin{eqnarray}
&& \left[ {E_\alpha ^a \left( { \rho _{A|B_1B_2\cdots B_{N-1} }} \right)} \right]^\mu   = f_\alpha ^\mu \left( {C\left( {\rho _{A|B_1B_2\cdots B_{N-1} } } \right)} \right) \le f_\alpha ^\mu  \left( 1 \right) \nonumber\\
  &=& f_\alpha ^\mu \left( {\sqrt {\left[ {C^a \left( {\rho _{A|B_1 } } \right)} \right]^2  +  \cdots  + \left[ {C^a \left( {\rho _{A|B_{k + 1} } } \right)} \right]^2  - T} } \right) \nonumber\\
  &\le& f_\alpha ^\mu \left( {\sqrt {\left[ {C \left( {\rho _{A|B_1 } } \right)} \right]^2  +  \cdots  + \left[ {C \left( {\rho _{A|B_{k + 1} } } \right)} \right]^2} } \right),
\end{eqnarray}
where we have used the monotonically increasing property of $f_\alpha(x)$ in the second inequality, in the forth inequality we have used (23) and the monotonically increasing property of $f_\alpha(x)$.

When $k+1\le m$, we have
\begin{eqnarray}
&& f_\alpha ^\mu \left( {\sqrt {\left[ {C \left( {\rho _{A|B_1 } } \right)} \right]^2  +  \cdots  + \left[ {C \left( {\rho _{A|B_{k + 1} } } \right)} \right]^2} } \right)\nonumber\\
  &\le& f_\alpha^\mu \left( {C \left( {\rho _{A|B_1 } } \right)} \right) + (2^{\mu}-1)f_\alpha^\mu \left( {C \left( {\rho _{A|B_2 } } \right)} \right) + \cdots  + (2^{\mu}-1)^k f_\alpha  \left( {C \left( {\rho _{A|B_{k + 1} } } \right)} \right) \nonumber\\
  &\le&  \left[ {E_\alpha ^a \left( {\rho _{A|B_1}}  \right)} \right]^\mu +(2^{\mu}-1)\left[ {E_\alpha ^a \left( {\rho _{A|B_2}}  \right)} \right]^\mu +\cdots+(2^{\mu}-1)^{m-1}\left[ {E_\alpha ^a \left( {\rho _{A|B_m}}  \right)} \right]^\mu \nonumber\\
&&~~~~~~~~~+(2^{\mu}-1)^{m+1}\left(\left[ {E_\alpha ^a \left( {\rho _{A|B_{m+1}}}  \right)} \right]^\mu +\cdots+\left[ {E_\alpha ^a \left( {\rho _{A|B_{N-2}}}  \right)} \right]^\mu \right)\nonumber\\
&&~~~~~~~~~+(2^{\mu}-1)^{m}\left[ {E_\alpha ^a \left( {\rho _{A|B_{N-1}}}  \right)} \right]^\mu,
\end{eqnarray}
where we have used (25) repeatedly and the similar consideration in the proof of Theorem 1 in the first inequality, and the second inequality is due to (19) and (23).

\begin{figure}
  \centering
  \includegraphics[width=7cm]{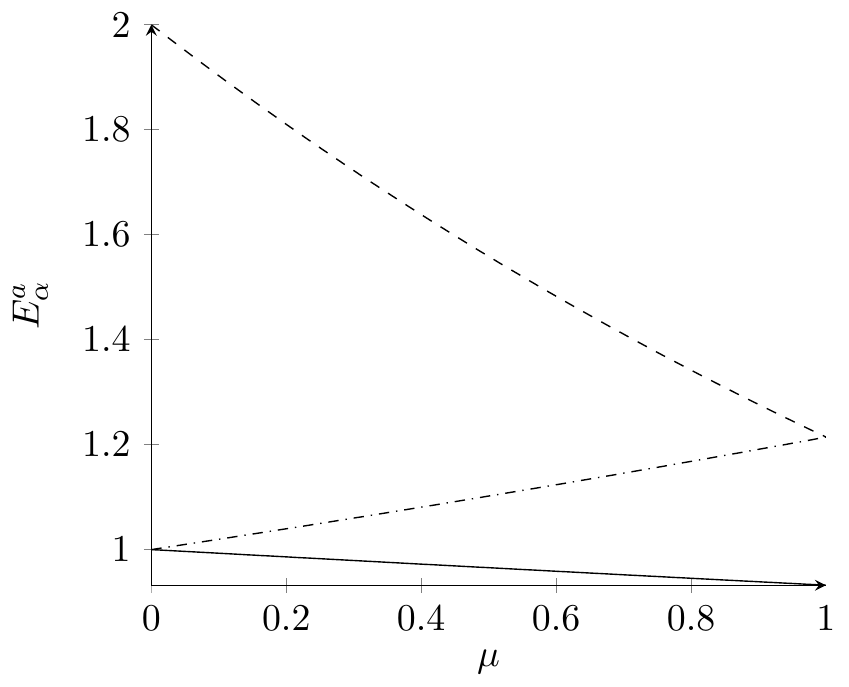}\\
  \caption{Behavior of the R\'{e}nyi-$\alpha$ entanglement of assistance of $|W\rangle$ and its upper bound, which are functions of $\mu$ plotted. The solid line represents the R\'{e}nyi-$\alpha$ entanglement of assistance of $|W\rangle$ in Example 2, the dot-dashed line represents the upper bound from our result, and the dashed line represents the upper bound from (21) in \cite{WS2}.}
\end{figure}

When $k+1 > m$, we have
\begin{eqnarray}
&& f_\alpha ^\mu \left( {\sqrt {\left[ {C \left( {\rho _{A|B_1 } } \right)} \right]^2  +  \cdots  + \left[ {C \left( {\rho _{A|B_{k + 1} } } \right)} \right]^2} } \right)\nonumber\\
  &\le& f_\alpha^\mu \left( {C \left( {\rho _{A|B_1} } \right)} \right) +(2^{\mu}-1)f_\alpha^\mu \left( {C \left( {\rho _{A|B_2} } \right)} \right)+ \cdots+(2^{\mu}-1)^{m-1}f_\alpha^\mu \left( {C \left( {\rho _{A|B_m} } \right)} \right) \nonumber\\
  && +(2^{\mu}-1)^{m+1}\left(f_\alpha^\mu \left( {C \left( {\rho _{A|B_{m+1}} } \right)} \right)+\cdots+f_\alpha^\mu \left( {C \left( {\rho _{A|B_{k}} } \right)} \right) \right)\nonumber\\
  && +(2^{\mu}-1)^{m} {f_\alpha ^\mu \left( C \left({\rho _{A|B_{k+1}}} \right) \right)} \nonumber\\
  &\le&  \left[ {E_\alpha ^a \left( {\rho _{A|B_1}}  \right)} \right]^\mu +(2^{\mu}-1)\left[ {E_\alpha ^a \left( {\rho _{A|B_2}}  \right)} \right]^\mu +\cdots+(2^{\mu}-1)^{m-1}\left[ {E_\alpha ^a \left( {\rho _{A|B_m}}  \right)} \right]^\mu \nonumber\\
&&~~~~~~~~~+(2^{\mu}-1)^{m+1}\left(\left[ {E_\alpha ^a \left( {\rho _{A|B_{m+1}}}  \right)} \right]^\mu +\cdots+\left[ {E_\alpha ^a \left( {\rho _{A|B_{N-2}}}  \right)} \right]^\mu \right)\nonumber\\
&&~~~~~~~~~+(2^{\mu}-1)^{m}\left[ {E_\alpha ^a \left( {\rho _{A|B_{N-1}}}  \right)} \right]^\mu,
\end{eqnarray}
where we have used (25) repeatedly and the similar consideration of the proof of Theorem  1 in the first inequality, and the second inequality is due to (19) and (23).

Combing (28), (29) and (30), we have completed the proof of Theorem 3.
\qed

Moreover, for the case that ${C_{AB_i}}\geqslant {C_{A|B_{i+1}\cdots B_{N-1}}}$ for all $i=1, 2, \cdots, N-2$, we have a simple tighter monogamy relation for the R\'{e}nyi-$\alpha$ entanglement of assistance:

\begin{theorem}
If ${C_{AB_i}}\geqslant {C_{A|B_{i+1}\cdots B_{N-1}}}$ for all $i=1, 2, \cdots, N-2$, we have
\begin{eqnarray}
&&~~~\left[ {E_\alpha ^a \left( {\rho _{A|B_1B_2\cdots B_{N-1} }} \right)} \right]^\mu  \nonumber\\
&&~~~~~~\leqslant \left[ {E_\alpha ^a \left( {\rho _{A|B_1}}  \right)} \right]^\mu +(2^{\mu}-1)\left[ {E_\alpha ^a \left( {\rho _{A|B_2}}  \right)} \right]^\mu+ \cdots+(2^{\mu}-1)^{N-2}\left[ {E_\alpha ^a \left( {\rho _{A|B_{N-1}}}  \right)} \right]^\mu  ,
\end{eqnarray}
for $0\leq\mu\leq1$ and $\alpha\in[(\sqrt{7}-1)/2,(\sqrt{13}-1)/2]$.
\end{theorem}

 As an example, let us consider the $W$ state, $|W\rangle=\frac{1}{\sqrt{3}}(|100\rangle+|010\rangle+|001\rangle).$ Set $\alpha=\left( {\sqrt 7  - 1} \right)/2\approx0.823$, then we have
$E_\alpha^a({|W\rangle_{A|B}})=E_\alpha^a({|W\rangle_{A|C
}})=0.607218$, $E_\alpha^a({|W\rangle_{A|BC}})=0.932108$, and then $\left[E_\alpha^a({|W\rangle_{A|BC}})\right]^{\mu}=(0.932108)^{\mu}$, $\left[E_\alpha^a({|W\rangle_{A|B}})\right]^{\mu}+\left[E_\alpha({|W\rangle_{A|C}})\right]^{\mu}=2(0.607218)^{\mu}$, $\left[E_\alpha^a({|W\rangle_{A|B}})\right]^{\mu}+\left[(2^{\mu}-1)E_\alpha^a({|W\rangle_{A|C}})\right]^{\mu}=2^{\mu}(0.607218)^{\mu}$ for $0\leq\mu\leq1$. It is easily verified that our results are better than the results in (21) for $0\leq\mu\leq1$; see Fig 2.

\section{conclusion}

Entanglement monogamy and polygamy relations are not only fundamental property of entanglement in multi-party systems but also provide us an efficient way of characterizing multipartite entanglement. We have presented monogamy relations satisfied by the $\mu$-th power of  R\'{e}nyi-$\alpha$ entanglement for $\mu\geqslant2$ and $\alpha\in[(\sqrt{7}-1)/2,(\sqrt{13}-1)/2]$, and polygamy relations satisfied by the $\mu$-th power of  R\'{e}nyi-$\alpha$ entanglement of assistance for $0\leq\mu\leq1$ and $\alpha\in[(\sqrt{7}-1)/2,(\sqrt{13}-1)/2]$. They are tighter , at least for some classes of quantum states, than the existing entanglement monogamy and polygamy relations. Tighter monogamy and polygamy relations imply finer characterizations of the entanglement distribution. Our approach may also be used to further study the monogamy and polygamy properties related to other quantum correlations.

\bigskip
\noindent{\bf Acknowledgments}\, \, This work is supported by the NSFC 11571119 and NSFC 11475178.

\end{document}